\documentstyle[12pt]{article}

\def\beq{\begin{equation}}
\def\eeq{\end{equation}}

\newcommand{\sect}[1]{\setcounter{equation}{0}\section{#1}}

\newcommand{\EQ}{\begin{equation}}
\newcommand{\EN}{\end{equation}}
\newcommand{\bea}{\begin{eqnarray}}
\newcommand{\ena}{\end{eqnarray}}

\begin{document}
%\begin{titlepage}
%\rightline{DFTT 41/96}
\rightline{NORDITA-99/14 HE}
\rightline{\hfill February 1999}
%rightline{hep-th/.....}
\vskip 1.2cm

\centerline{\Large \bf An introduction to AdS/CFT equivalence}
%\centerline{\Large \bf of all interactions?}

\vskip 1.2cm

\centerline{\bf Paolo Di Vecchia}
% and Lorenzo Magnea\footnote{On leave from Universit\`a di Torino, Italy}}
\centerline{\sl NORDITA}
\centerline{\sl Blegdamsvej 17, DK-2100 Copenhagen \O, Denmark}
\vspace{1cm}
\centerline{ {\it Talk given at the 32nd International Symposium on the 
Theory of Elementary Particles,}}
\centerline{\it{Buckow, Germany  (September 1998)}}
%\vspace{1cm}
%\vskip .3cm
%
%\vskip .2cm
%
%\centerline{\bf Alberto Lerda\footnote{II Facolt\`a di Scienze
%M.F.N., Universit\`a di Torino
%(sede di Alessandria), Italy}}
%\centerline{\sl Dipartimento di Scienze e Tecnologie Avanzate and}
%\centerline{\sl Dipartimento di Fisica Teorica, Universit\`a di Torino}
%\centerline{\sl Via P.Giuria 1, I-10125 Torino, Italy}
%\centerline{\sl and I.N.F.N., Sezione di Torino}
%
%\vskip .2cm
%
%\centerline{\bf Raffaele Marotta\footnote{Della Riccia fellow}}
%\centerline{\sl Dipartimento di Scienze Fisiche, Universit\`a di Napoli}
%\centerline{\sl Mostra D'Oltremare, Pad. 19, I-80125 Napoli, Italy}
%
%\vskip .2cm
%
%\centerline{\bf Rodolfo Russo}
%\centerline{\sl Dipartimento di Fisica, Politecnico di Torino}
%\centerline{\sl Corso Duca degli Abruzzi 24, I-10129 Torino, Italy}
%\centerline{\sl and I.N.F.N., Sezione di Torino}
%\vskip 1cm

\begin{abstract}
This is an introduction to the Maldacena conjecture on the
equivalence between ${\cal{N}}=4$ super Yang-Mills in Minkowski space-time 
and type IIB string theory compactified on $AdS_5 \otimes S_5 $.
\end{abstract}

%\end{titlepage}
%\end{document}
\vskip 1cm
%\newpage
%\renewcommand{\thefootnote}{\arabic{footnote}}
%\setcounter{footnote}{0}
%\setcounter{page}{1}
%
\sect{Introduction}
\label{intro}
%\vskip 0.5cm

Gravitational interactions are described by the Einstein's theory of
general relativity or more in general by supergravity that is its 
supersymmetric 
extension, while the other interactions are described by  gauge field 
theories. Actually also the theory of general relativity is a gauge theory
corresponding to the gauging of the space-time Poincar{\'{e}} group, while
those that are more usually called gauge theories correspond to the
gauging of an internal symmetry group. But apart from the fact that they are
both gauge theories does it exist any deeper relation between them? Do they
imply each other in a consistent quantum theory of gravity? 
In the framework of field theory there is no connection; they 
can both exist independently from each 
other, but any field theory involving gravity suffers from the problem of
non-renormalizability. However, when we try to look at this question in
the framework of string theory, we see that they occur naturally
together in the same theory and actually it has not  been possible to 
construct a string theory without both of them.

String theories were born from the attempt of describing the properties of
the strong interactions through the construction of the dual resonance model.
It became soon clear, however, that this model in its consistent form, 
that later on was recognized to correspond to the quantization of  a 
relativistic string,
contained all sort of massless particles as gluons, gravitons and others
except a massless pseudoscalar particle corresponding to the pion that in the
chiral limit is
the only massless particle that we expect in strong interactions. 
Because of this and other unphysical features 
it became clear since the middle of the seventies that string theories could 
not provide a theory for strong
interactions, that in the meantime were successfully described in the
framework of QCD, but could instead be used as a consistent way of unifying all
interactions in a theory containing also quantum gravity~\cite{SCHERK}. 
It turns out  that all five consistent
string theories in ten dimensions all unify in a way or another gravity
with gauge theories. Let us remind now how this comes about. 

The type I theory is a theory of open and closed string.
Open strings have Chan-Paton gauge degrees of freedom located at the 
end points and, because of this, an open string theory contains the usual
gauge theories. On the other hand a pure theory of open strings is not
consistent by itself; non-planar loop corrections generate closed strings
and a closed string theory contains gravity. Therefore in the type I
theory open strings require for consistency  closed strings. This implies that
gravity, that is obtained in the zero slope limit 
of closed strings, is a necessary consequence of gauge 
theories, that are obtained in the zero slope limit of the open string theory.
The heterotic strings is instead a theory of only closed strings that 
contains, however, both supergravity and gauge theories. But in this case 
gravity is the fundamental theory and gauge theories are
obtained from it through a stringy Kaluza-Klein mechanism. The remaining 
consistent
theories in ten dimensions are the two type II theories that at the 
perturbative level contain only closed strings and no gauge degrees of freedom. 
However, they also contain non-perturbative objects, the $D$-branes that are 
characterized by the fact that open strings can end on them. Therefore
through the D-branes open strings also appear in type II theories and
with them we get also gauge theories.
In conclusion all string theories contain both gravity and gauge theories
and therefore those two kinds of interactions are intrinsically unified
in string theories. But, since all string theories contain
gravity, it seems impossible to use a string theory to describe strong
interactions. In fact they are described by QCD that does not
contain gravity!!

On the other hand it is known since the middle of the seventies that, if we
consider a non-abelian gauge theory with gauge group $SU(N)$ and we take
the 't Hooft limit where the number of colours $N \rightarrow \infty$,
while the product $g^{2}_{YM} N \equiv \lambda$ is kept fixed~\cite{HOOFT}, 
the gauge theory simplifies in the sense that the only diagrams
surviving in this limit are the planar ones. In the large $N$ limit it can be
shown that the gauge invariant observables are determined by a master
field~\cite{WITTEN1} that satisfies a classical equation of motion. It has 
also been
conjectured that in this limit QCD is described by a string theory; the
mesons are string excitations that are free when $N \rightarrow \infty$.
This idea is also supported by the experimental fact that hadrons lie on 
linearly rising 
Regge trajectories as required by a string model. The fact that the large
$N$ expansion may be a good approximation also for low values of $N$ as
$N=3$ in the case of QCD is suggested by the validity of the Zweig's rule and 
by the successful  explanation of the $U(1)$-problem in the 
framework of the large $N$ expansion~\footnote{For an early  review see for 
instance Ref.~\cite{PAOLO}.}. The fact, however, that 
any consistent 
string theories includes necessarily gravity has led to call the
string theory coming out from QCD as the QCD string because, as QCD, it should
not contain gravity. Although many attempts
have been made to construct a QCD string none can be considered sufficiently
satisfactory. This problem has been with us for the last thirty years.
 
A more recent connection between gauge theories and gravity comes from the
D-branes.
A system of $N$ coincident D $p$-branes is a classical solution of the 
low-energy
string effective action in which only the metric, the dilaton and a RR 
$(p+1)$-form potential are different from zero. The metric is given by:
\beq 
(ds)^2 = H^{-1/2} (y) \eta_{\alpha \beta} d x^{\alpha} d x^{\beta} + H^{1/2} (y)
\delta_{ij}d y^{i} d y^{j}
\label{clasol}
\eeq
while the dilaton and RR potential are equal to:
\beq
e^{- (\phi - \phi_0) } = [ H(y) ]^{(p-3)/4} \hspace{1cm}; \hspace{1cm} A_{01 
\dots p}= [H(y)]^{-1}
\label{clasol2}
\eeq
where
\beq
H (y) = 1 + \frac{K_p N}{r^{7-p}} \hspace{2cm} K_p = 
\frac{(2 \pi \sqrt{\alpha '})^{7-p}}{(7-p) \Omega_{8-p}} g_s
\label{acca}
\eeq
with $r^2 \equiv y_i y^i$ and $\Omega_{q} = 2 \pi^{(q+1)/2}/\Gamma[(q+1)/2]$.
The indices $\alpha $ and $\beta$ run along the world volume of the brane, while
the indices $i$ and $j$ run along the directions that are transverse to the 
brane.

%At low energy a single D $p$-brane is described by the abelian Born-Infeld
%action:
%\beq
%S_{BI} = - \tau_{p}^{(0)} \int d^{p+1} \xi e^{-\phi} \sqrt{- \det \left[ 
%G_{\alpha \beta } + 2 \pi \alpha ' F_{\alpha \beta} \right]}
%\label{bi}
%\eeq

A system of $N$ coincident D $p$-branes is described by the non-abelian version
of the Born-Infeld action, whose complete form is not yet known, but
for our considerations we can take of the form suggested in Ref.~\cite{TSE}:
\beq
S_{BI} = - \tau_{p}^{(0)} \int d^{p+1} \xi \,\, e^{-\phi} STr \,\, \sqrt{- \det 
\left[ G_{\alpha \beta } + 2 \pi \alpha ' F_{\alpha \beta} \right]}
\label{bi3}
\eeq
The brane tension is given by:
\beq
\tau_p \equiv \frac{\tau_{p}^{(0)}}{g_s} = 
\frac{(2 \pi \sqrt{\alpha'})^{1-p}}{2 \pi \alpha' g_s} \hspace{2cm} g_s \equiv
e^{\phi_0}
\label{pten}
\eeq 
where the string coupling constant $g_s$ is identified with the value at 
infinity of the dilaton field. $G_{\alpha \beta}$ is the pullback of the metric 
$G_{\mu \nu}$ and $F_{\alpha \beta}$ is a gauge field leaving on the brane.
$S Tr $ stands for a symmetrized trace over the group matrices. 
By expanding the Born-Infeld action in powers
of $F_{\alpha \beta}$ we find at the second order the kinetic term
for a non abelian gauge field (the $U(N)$ matrices are normalized as 
$Tr (T_i T_j ) = \frac{1}{2} \delta_{ij}$):
\beq
S_{BI} = - \frac{1}{4 g_{YM}^{2}} \int d^{p+1} \xi \,\, F^{a}_{\mu \nu} 
F^{a \mu \nu} \hspace{.5cm};\hspace{.5cm} 
g^{2}_{YM} = 2 g_s (2 \pi)^{p-2} (\alpha ')^{(p-3)/2}
\label{biexp}
\eeq
%where
%\beq
%\label{YMcc}
%\eeq
We see here another connection between gravity and gauge theories. On one hand 
the D-branes are classical solutions of the low-energy supergravity action, 
while on the other hand they are described
by a gauge field theory whose action reduces at low-energy to the usual
action of Yang-Mills theory.
  
An interesting property of the D-brane solution in eqs.(\ref{clasol}) and 
(\ref{clasol2}) is that for large values of $r$ the metric  becomes flat.
Therefore, being the curvature small, the classical supergravity description
provides a good approximation of the D brane. In the following we
want to study the behaviour of the classical solution in the near-horizon limit
corresponding to $r \rightarrow 0$ for $p=3$ for which the dilaton in 
eq.(\ref{clasol2}) is independent of $r$ and the Yang-Mills coupling constant 
in eq.(\ref{biexp}) is dimensionless implying that the four-dimensional world
volume theory is conformal invariant. More precisely the near-horizon limit
is defined by:
\beq
r \rightarrow 0 \hspace{2cm} \alpha ' \rightarrow 0 \hspace{2cm} U \equiv
\frac{r}{\alpha '} = fixed
\label{lim}
\eeq
in which also the Regge slope is taken to zero, while $U$ is kept fixed. In 
this limit we can neglect the factor $1$ in the function $H$ in 
eq.(\ref{acca}) and the metric in eq.(\ref{clasol}) becomes:
\beq
\frac{(ds)^2}{\alpha'} \rightarrow \frac{U^2}{\sqrt{4 \pi N g_{s}}} (dx_{3+1})^2
+ \frac{\sqrt{4 \pi N g_{s}}}{U^2} dU^2 + \sqrt{4 \pi N g_{s}} d \Omega_{5}^{2}
\label{nmet}
\eeq
This is the metric of the manifold $AdS_5 \times S_5 $ where the two radii of
$AdS_5$ and $S_5$ are equal and given by:
\beq
R_{AdS_5}^{2} = R_{S_5}^{2} \equiv b^2 = \alpha' \sqrt{4 \pi N g_{s}} 
\label{rad}
\eeq
that, using the relation $g_{YM}^{2} = 4 \pi g_s$ following from 
eq.(\ref{biexp}), implies:
\beq
\frac{b^2}{\alpha '} = \sqrt{N g_{YM}^{2}}
\label{rad2}
\eeq
The world volume theory of $N$ coincident D $3$-branes is ${\cal{N}} =4$
super Yang-Mills theory in $3+1$ dimensions with $U(N)$ gauge group. On
the other hand the classical solution in eq.(\ref{nmet}) is a good 
approximation when the radii of $AdS_5$ and $S_5$ are very big:
\beq
\frac{b^2}{\alpha'} >> 1 \Longrightarrow Ng_{YM}^{2} \equiv \lambda >>1
\label{bigrad}
\eeq
The fact that those two descriptions are simultaneously consistent for 
large values of the coupling constant $\lambda$ brought Maldacena~\cite{MALDA} 
to formulate the conjecture that the strongly interacting ($\lambda >>1$)
${\cal{N}}=4$ super Yang-Mills with gauge group $U(N)$ is actually equivalent
to the ten-dimensional classical supergravity compactified on 
$AdS_5 \times S_5$. 
But supergravity is not a consistent quantum theory
and therefore in order to extend the conjecture to any value of $\lambda$ one 
has to
find a substitute for classical supergravity. The natural way of extending the
previous conjecture is therefore to  say that  ${\cal{N}}=4$ super Yang-Mills 
is equivalent to type IIB string theory compactified on the special background
$AdS_5 \times S^5$~\cite{MALDA}. The parameters of the Yang-Mills theory 
$g_{YM}^{2}$ and 
$N$  are related to the parameters of string theory $g_s$, $R_{S_5}$ and 
$\alpha '$ through the following relations:
\beq
g^{2}_{YM} \equiv \frac{\lambda}{N} = 4 \pi g_s \hspace{2cm}
R_{S_5}^{2} = R^{2}_{AdS_5} = \lambda^{1/2} \alpha '
\label{relpar}
\eeq 
In conclusion, according to the Maldacena conjecture, classical supergravity
is a good approximation if $\lambda >>1$, while in 't Hooft limit in which
$\lambda$ is kept fixed for $N \rightarrow \infty$ classical string theory
is a good approximation for  ${\cal{N}}=4$ super Yang-Mills. In the 't Hooft
limit in fact string loop corrections are negligible ($g_s << 1$) 
as one can see from the first eq. in (\ref{relpar}). Finally Yang-Mills
perturbation theory is a good approximation when $\lambda <<1$.
The strongest evidence for the validity of the Maldacena conjecture comes from
the fact that both ${\cal{N}}=4$ super Yang-Mills and type IIB string 
compactified on $AdS_5 \otimes S_5$ have the same symmetries. They are, in 
fact, both invariant under $32$ supersymmetries, under the conformal group
$O(4,2)$, corresponding to the isometries of $AdS_5$, under the $R$-symmetry
group $SU(4)$, corresponding to the isometries of $S_5$ and under the 
Montonen-Olive
duality~\cite{MONTOLI} based on the group $SL(2,Z)$. It is important to stress
that the two theories live on different spaces: IIB string theory lives on 
$AdS_5 \otimes S_5$, while ${\cal{N}} =4$ super Yang-Mills lives on the
boundary of $AdS_5$ that is our four-dimensional Minkowski space. This is
an explicit realization that a normal four-dimensional Yang-Mills theory, as
also QCD is, can be described by a string theory without running into the 
problem that a string theory contains gravity while a gauge theory 
does not. This old problem is solved in this case by the fact that
the gauge and the string theories live in different spaces. A new puzzle,
however, arises in this case because we usually connect a string theory
with a confining gauge theory, while instead ${\cal{N}}=4$ super Yang-Mills
is a conformal invariant theory and therefore is in the Coulomb and not in
the confining phase.
%\section{Correspondence between fields}
\par
But apart from this puzzle, if two theories, as the type IIB string theory 
compactified on 
$AdS_5 \otimes S_5$ and ${\cal{N}}=4 $ super Yang-Mills theory, are 
equivalent then it must be possible to specify for each field $Q(x)$
of the boundary Minkowski theory the corresponding field $\Phi(y)$ of the bulk
string theory 
and to show that, when we compute corresponding correlators in the two theories,
we get the same result. In particular, in the boundary theory one can
easily compute the generating functional for correlators involving  $Q(x)$

\beq
Z(\Phi_0 ) = < e^{\int d^4 x \Phi_0 (x) Q(x)} >
\label{gene1}
\eeq
By taking derivatives with respect to the arbitrary source
$\Phi_0 (x)$ one can compute any correlator involving the boundary field
$Q(x)$. In Refs.~\cite{KLEBA,WITTEN2} the recipy for computing $Z(\Phi_0)$
in the bulk theory has been given. First of all 
one must identify $\Phi_0 (x)$ with the boundary value of the field
$\Phi (y)$, which lives in the bulk theory and that corresponds to 
$Q(x)$ of the boundary theory. Then the generating functional given in
eq.(\ref{gene1}) is just obtained by performing in the bulk theory
the functional integral
over $\Phi$ with the restriction that its boundary value be $\Phi_0$:
\beq
Z( \Phi_0 ) = \int_{\Phi \rightarrow \Phi_0} D \Phi e^{-S [ \Phi]}
\label{gene2}
\eeq
In computing the previous functional integral we can use classical 
supergravity in the regime where $\lambda >> 1$. Otherwise for an
arbitrary value of fixed $\lambda$ for $N \rightarrow \infty$ we need to
compute
the tree diagrams of type IIB string theory compactified on $AdS_5 \otimes
S_5$. 

A number of bulk fields have been identified to correspond to the various
gauge invariant composite fields of ${\cal{N}}=4$ super Yang-Mills. We
do not have the time in this talk to describe them in detail. In the
following we will just describe in some detail the correspondence between
the dilaton field of type IIB supergravity and the composite given by the
Yang-Mills Lagrangian $F^2 \equiv F_{\mu \nu}^{a} F^{a \mu \nu}$ showing
in detail that the two-point functions that one obtains from both 
eqs.(\ref{gene1}) and (\ref{gene2})  are  coincident~\cite{KLEBA,WITTEN2}.

Since ${\cal{N}}=4$ super Yang-Mills theory with gauge group $SU(N)$ is a
conformal invariant quantum theory and the composite field $F^2$ has
dimension $4$ the two-point function involving two $F^2$ fields must have 
the following form:
\beq
< F^2 (x) F^2 (z) >  \sim \frac{N^2}{(\vec{x} -\vec{z})^2}
\label{f2f2}
\eeq    
apart from an overall constant that we do not care to compute. The previous
correlator can also be obtained by using the lowest order perturbation
theory in ${\cal{N}}=4$ super Yang-Mills. $\vec{x}$ denotes here a Minkowski
four-vector.

In the bulk theory we only need the dilaton kinetic term in
type IIB supergravity in $D=10$ compactified on $AdS_5 \otimes S_5$. Taking
into account that the volume of $S_5$ is equal to $\pi^3 b^3$, where $b$ is
given in eq.(\ref{rad}), we need to consider the following action:
\beq
S =\frac{\pi^3 b^3}{4 \kappa^{2}_{10}} \int d^5 x \sqrt{g} g^{\mu \nu}
\partial_{\mu} \Phi \partial_{\nu} \Phi 
\label{act}
\eeq
where $g_{\mu \nu} = \frac{b^2}{x_{0}^{2}} \delta_{\mu \nu}$ is the metric
of $AdS_5$ in the so-called Poincar{\'{e}} coordinates with $\mu, \nu =0 
\dots 4$. In the limit $\lambda >> 1$, where classical supergravity is a good
approximation, we just need to solve the dilaton eq. of motion given by:
\beq
\partial_{\mu} \left[ \sqrt{g} g^{\mu \nu} \partial_{\nu} \Phi \right] =0
\label{eqmo}
\eeq
The solution of the previous equation, that is equal to $\Phi_0$ on the
boundary (corresponding to the limit $x_0 \rightarrow 0$), can be given in 
terms of the Green's function:
\beq
\Phi (x_0, \vec{x}) = \int d^4 {\vec{x}}\,\, K( x_0 , \vec{x}; \vec{z}  )
\,\,\Phi_0 ( \vec{z}  )
\label{solo2} 
\hspace{.5cm};\hspace{.5cm}
K( x_0 , \vec{x}; \vec{z}  ) \sim \frac{x_{0}^{4}}{[x_{0}^{2} + ( \vec{x}
- \vec{z}  )^2 ]^4}
\eeq
Inserting the solution found in eq.(\ref{solo2}) in the classical action we get
that the contribution to the classical action is entirely due to the
boundary term
\beq
S = \frac{\pi^3 b^8}{4 \kappa_{10}^{2}} \int d^4 \vec{x} x_{0}^{-3} \Phi
\partial_0 \Phi |_{\epsilon}^{\infty} \sim - \frac{\pi^3 b^8}{4 
\kappa_{10}^{2}} \int d^4 \vec{x}\int d^4 \vec{z}
\frac{\Phi_0 (\vec{x}) \Phi_0 (\vec{z} ) }{(\vec{x} - \vec{z}  )^8 }
\label{claact}
\eeq 
where we have introduced a cut off $\epsilon$ at the lower limit of integration,
that, however, cancels out after having inserted  eq.(\ref{solo2}) in
eq.(\ref{claact}).
In conclusion in the classical approximation ($\lambda >>1$) we get
\beq
Z ( \Phi_0 ) = \exp \left[\frac{\pi^3 b^8}{4 \kappa_{10}^{2}} \int d^4 \vec{x}
\int d^4 \vec{x}'
\frac{\Phi_0 (\vec{x}) \Phi_0 (\vec{x}' ) }{(\vec{x} - \vec{x} ' )^8 } \right]
\label{clagen}
\eeq
Taking into account eq.(\ref{rad}) and that $2 \kappa_{10}^{2} = ( 2 \pi)^7 
g_{s}^{2} (\alpha ')^4 $, from the previous equation we can get
immediately the two-point function:
\beq
< F^2 (x) F^2 (z) > =
\frac{\partial^2 Z (\Phi_0 )}{\partial \Phi_0 (\vec{x} ) \partial \Phi_0 
(\vec{z} ) }  \sim \frac{N^2}{(\vec{x} - \vec{z} )^8} 
\label{corre}
\eeq 
that agrees with the expression given in eq.(\ref{f2f2}).

\vskip0.5cm
\noindent
{\large \bf Acknowledgements}

\smallskip
\noindent
I thank J.L. Petersen for a continous exchange of information about the
AdS/CFT correspondence and M. Frau, A. Lerda and R. Russo for
many useful discussions.

%%%%%%%%%%%%%%%%%%%%%%
%%%%%%%%%%%%%%%%%%%%%%

\end{document}